\documentclass[aps,prd,preprint,superscriptaddress,tightenlines,nofootinbib]{revtex4}

\usepackage{graphicx}% Include figure files
\usepackage{dcolumn}% Align table columns on decimal point
\usepackage{bm}% bold math

\begin{document}

\preprint{CLNS 04-1887}       % for CLNS notes
\preprint{CLEO 04-11}         % for CLNS notes

\title{Measurement of the Muonic Branching Fractions \\
of the Narrow Upsilon Resonances}

\author{G.~S.~Adams}
\author{M.~Chasse}
\author{M.~Cravey}
\author{J.~P.~Cummings}
\author{I.~Danko}
\author{J.~Napolitano}
\affiliation{Rensselaer Polytechnic Institute, Troy, New York 12180}
\author{D.~Cronin-Hennessy}
\author{C.~S.~Park}
\author{W.~Park}
\author{J.~B.~Thayer}
\author{E.~H.~Thorndike}
\affiliation{University of Rochester, Rochester, New York 14627}
% ...
\collaboration{CLEO Collaboration}
\noaffiliation

\date{September 3, 2004}

\begin{abstract}
The decay branching fractions of the three narrow $\Upsilon$ resonances to $\mu^+ \mu^-$ have been measured by analyzing about 4.3 fb$^{-1}$ $e^+e^-$ data collected with the CLEO III detector. The branching fraction ${\cal B}(\Upsilon(1S) \to \mu^+ \mu^-) = (2.49 \pm 0.02 \pm 0.07)\%$ is consistent with the current world average but ${\cal B}(\Upsilon(2S) \to \mu^+ \mu^-) = (2.03 \pm 0.03 \pm 0.08)\%$ and ${\cal B}(\Upsilon(3S) \to \mu^+ \mu^-) = (2.39 \pm 0.07 \pm 0.10)\%$ are significantly larger than prior results. These new muonic branching fractions imply a narrower total decay width for the $\Upsilon$(2S) and $\Upsilon$(3S) resonances and lower other branching fractions that rely on these decays in their determination.
\end{abstract}

\maketitle

Recent advances in Lattice QCD promise accurate predictions for a wide variety of non-perturbative quantities \cite{preciseLQCD}. 
However, substantially improved data are needed to confront these predictions.
The long-lived $b\bar{b}$ states are especially well suited for establishing the accuracy of Lattice QCD calculations \cite{LQCD} as well as testing effective theories of the strong interactions, such as potential models \cite{Yreview}, in the heavy quark sector.
The large data sample collected recently by the CLEO detector in the vicinity of the $\Upsilon$(nS) ($n=1, 2, 3$) resonances enables us to determine the $b\bar{b}$ resonance parameters, such as the leptonic and total decay widths, with unprecedented precision. 

The total widths of the narrow $\Upsilon$ resonances below the open-beauty threshold cannot be measured directly since their natural widths ($25-50$ keV) are much narrower than the collider beam energy spread ($4-5$ MeV). 
The common indirect method to determine the total decay width ($\Gamma$) is to combine the leptonic branching fraction (${\cal B}_{\ell\ell}$) with the leptonic decay width ($\Gamma_{\ell\ell}$): $\Gamma = \Gamma_{\ell\ell}/{\cal B}_{\ell\ell}$ \cite{Yreview,PDG}.
In practice, assuming lepton universality ($\Gamma_{ee} = \Gamma_{\mu\mu} = \Gamma_{\tau\tau}$), the leptonic decay width is replaced by $\Gamma_{ee}$, which can be extracted from the energy-integrated resonant hadron production cross section in $e^+e^-$ collisions, while the leptonic branching fraction is replaced by the muonic branching fraction, ${\cal B}_{\mu\mu} \equiv {\cal B}(\Upsilon \to \mu^+\mu^-$), which can be measured more accurately than ${\cal B}_{ee}$ or ${\cal B}_{\tau\tau}$. 
Therefore, it is very important to measure ${\cal B}_{\mu\mu}$ precisely in order to determine the total decay widths of the narrow $\Upsilon$ resonances.

The leptonic branching fraction is also interesting in its own right since it represents the strength of the $\Upsilon$ decay to lepton pairs via annihilation to a virtual photon.
Furthermore, ${\cal B}_{\mu\mu}$ is generally used in determinations of the branching fractions of hadronic and electromagnetic transitions among the $\Upsilon$ states since these decays are often measured by observing the decay of the lower lying resonances to lepton pairs.
In addition, comparing ${\cal B}_{\mu\mu}$ to ${\cal B}_{ee}$ as well as to ${\cal B}_{\tau\tau}$ can provide a check of lepton universality and test the possible existence of new physics beyond the Standard Model \cite{Higgs}. 

Based on previous measurements, ${\cal B}_{\mu\mu}$ has been established with a 2.4\% accuracy for the $\Upsilon$(1S) \cite{PDG}, and a modest 16\% and 9\% accuracy for the $\Upsilon$(2S) \cite{CLEO_2S,ARGUS,CUSB,CBAL}, and $\Upsilon$(3S) \cite{CUSB,CLEO_3S}, respectively.
This paper reports the measurement of ${\cal B}_{\mu\mu}$ for all three narrow resonances with a much larger data set and a more advanced detector. 
The new results enable us to determine the total decay widths of the $\Upsilon$(2S) and $\Upsilon$(3S) with better precision.

The data used in this analysis were collected by the CLEO~III detector at the Cornell Electron Storage Ring, a symmetric $e^+ e^-$ collider.
The analysis relies on the excellent charged particle tracking, electromagnetic calorimetry, and muon identification of CLEO~III.
The new tracking system consists of a 4-layer double-sided silicon vertex detector and a 47-layer drift chamber \cite{DR3} residing in a 1.5 T solenoidal magnetic field.
The crystal calorimeter and the muon detector system inherited from CLEO~II \cite{CLEOII} can identify muons with momentum above 1.0 GeV/$c$ with high efficiency.

To determine ${\cal B}_{\mu\mu}$, we measure 
$\tilde{\cal{B}}_{\mu\mu} \equiv \Gamma_{\mu\mu}/\Gamma_{\rm had}=(\tilde{N}_{\mu\mu}/\varepsilon_{\mu\mu})/(\tilde{N}_{\rm had}/\varepsilon_{\rm had})$, 
where $\Gamma_{\mu\mu}$ ($\Gamma_{\rm had}$) is the rate for $\Upsilon$ decay to $\mu^+\mu^-$ (hadrons), and $\tilde{N}$ and $\varepsilon$ are the number of observed (raw) resonance decays and the selection efficiency, respectively. 
$\Gamma_{\rm had}$ includes all decay modes of the resonances other than $e^+e^-$, $\mu^+\mu^-$, and $\tau^+\tau^-$.
Assuming lepton universality we have
${\cal B}_{\mu\mu} = \Gamma_{\mu\mu}/\Gamma = {\tilde{\cal{B}}}_{\mu\mu}/(1+3{\tilde{\cal{B}}}_{\mu\mu})$.

The major source of background is non-resonant (continuum) production of $\mu^+\mu^-$ and hadrons via $e^+ e^- \to \mu^+\mu^-$ and $e^+ e^- \to q\bar{q}$ ($q=u,d,c,s$), respectively, which cannot be distinguished experimentally from the corresponding resonance decays.
Hence, we use continuum data collected at energies just below each resonance  to subtract these backgrounds. 
The observed number of $\Upsilon$ decays to $\mu^+\mu^-$ (or hadron) is $\tilde{N} = \tilde{N}_{on} - S \tilde{N}_{off}$, where $S$ scales the luminosity of the off-resonance data to that of the on-resonance data and accounts for the $1/s$ dependence of the cross section.

Backgrounds from non-resonant $e^+e^- \to \tau^+\tau^-$, two-photon fusion ($e^+e^- \to e^+e^- \gamma^{\star}\gamma^{\star}$), or from radiative return to the lower resonances contribute less than 0.2\% after the off-resonance subtraction.
The remaining backgrounds (to $\mu^+\mu^-$) are mainly from cosmic rays, and more importantly from $\Upsilon$(2S) and $\Upsilon$(3S) decays to a lower $\Upsilon$ state, which decays to $\mu^+\mu^-$ and the accompanying particles escape detection. 
The background from $\Upsilon \to \tau^+\tau^-$ is negligible ($<0.05$\%) in the $\mu^+\mu^-$ measuremen but it is significant in the hadron measurement.

Our results are based upon 1.1 fb$^{-1}$ (1S), 1.2 fb$^{-1}$ (2S), and 1.2 fb$^{-1}$ (3S) data collected within 2-3 MeV at the peak of each resonance (``on-resonance samples'') as well as off-resonance samples which were collected 20-30 MeV below the resonances and represent 0.19 fb$^{-1}$ (below 1S), 0.44 fb$^{-1}$ (below 2S), and 0.16 fb$^{-1}$ (below 3S).
The scale factors, $S$, between the on-resonance and the corresponding off-resonance samples are calculated from the luminosity measured with the $e^+ e^- \to \gamma\gamma$ process \cite{lumi} which, unlike the $e^+ e^-$ final state, is not contaminated by resonance decays.

We select $\mu^+\mu^-$ events by requiring exactly two oppositely charged tracks, each with momentum between 70\% and 115\% of $E_{\rm beam}$, with polar angle $| \cos\theta | < 0.8$, and with the opening angle of the tracks greater than 170$^{\circ}$.
Muon identification requires each track to deposit $0.1-0.6$ GeV in the electromagnetic calorimeter, characteristic of a minimum ionizing particle, and at least one track to penetrate deeper than five interaction lengths into the muon chambers.

\begin{figure}
\includegraphics*[width=4.5in]{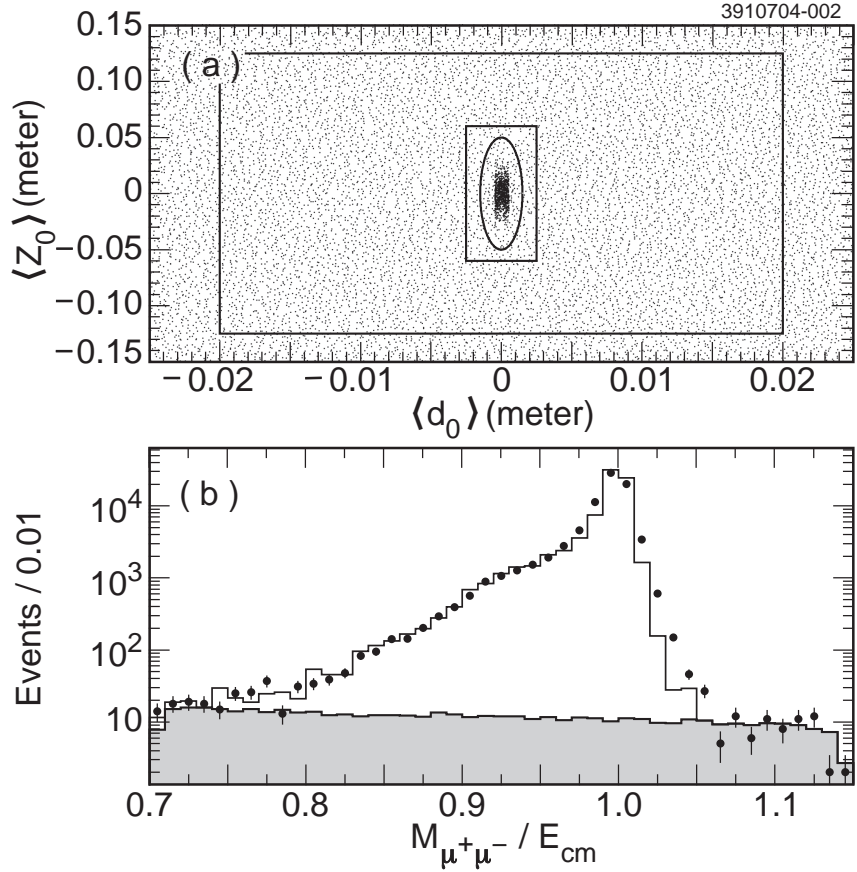}
\caption{(a) Distribution of $\mu^+\mu^-$ candidate events in off-resonance data below the $\Upsilon$(3S) over the $\langle z_0 \rangle$ vs. $\langle d_0 \rangle$ plane. The ellipse encircles the signal region while the two rectangles define the sideband. (b) Scaled invariant mass distribution of the $\mu^+\mu^-$ candidates in the signal region (dots) overlaid with the expected distribution from Monte Carlo simulation of $e^+e^- \to \mu^+\mu^-$ events. The shaded histogram represents the scaled distribution for events in the sideband. The vertical scale is logarithmic.}
\label{fig:cosmic}
\end{figure}

We control the cosmic-ray background using the track impact parameters with respect to the $e^+e^-$ interaction point (beam spot).
From the location of the point nearest to the beam spot (as seen in the plane perpendicular to the beam axis) on each track we calculate the separation between the two tracks along the beam axis ($\Delta z_0$) and in the perpendicular plane ($\Delta d_0$) as well as their average distance from the beam spot along the beam axis ($\langle z_0 \rangle$) and in the perpendicular plane ($\langle d_0 \rangle$).
We require $| \Delta z_0 | < 5$ cm, $| \Delta d_0 | < 2$ mm and
$(\langle z_0 \rangle/5\ {\rm cm})^2 + (\langle d_0 \rangle/1.5\ {\rm mm})^2 <1$.
Cosmic events are uniformly distributed in the $\langle z_0 \rangle$ and $\langle d_0 \rangle$ variables while events from $e^+e^-$ collisions populate a small area around $(\langle z_0 \rangle,\langle d_0 \rangle)=(0,0)$. 
We use a two-dimensional sideband in $\langle z_0 \rangle$ and $\langle d_0 \rangle$ to estimate the remaining cosmic-ray background (Fig.~\ref{fig:cosmic}a). This background is $0.3$\%$-0.6$\% depending on the data sample.
The observed rate of events with $M_{\mu\mu}>1.1E_{\rm cm}$ (after the momentum cuts have been relaxed) is consistent within 10\% with these background estimates (Fig.~\ref{fig:cosmic}b).
We correct the number of $\mu^+\mu^-$ events observed in the on-resonance and off-resonance samples individually for the cosmic background.

\begin{figure}[h]
\includegraphics*[width=4.5in]{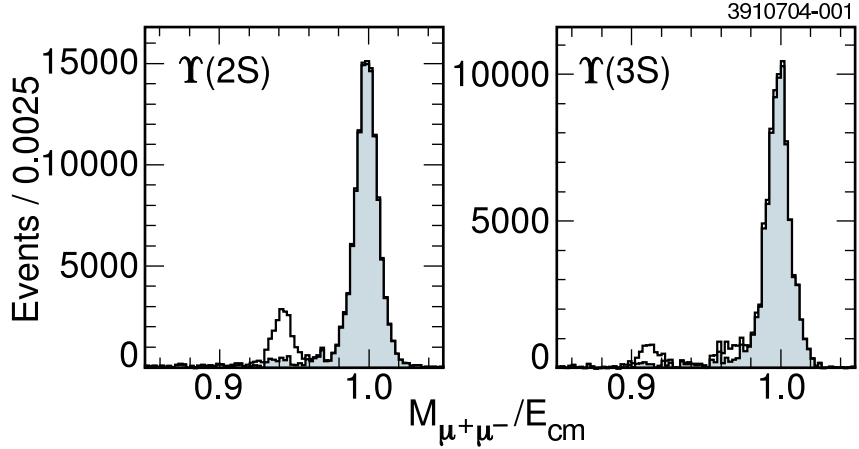}
\caption{Distribution of the scaled invariant mass of $\mu^+\mu^-$ candidates from the $\Upsilon$(2S) (left) and $\Upsilon$(3S) (right) after off-resonance subtraction. The empty (shaded) histograms show the distributions before (after) rejecting events with extra showers in the calorimeter.}
\label{fig:cascade}
\end{figure}

Requiring exactly two tracks suppresses the indirect $\mu^+\mu^-$ production at the $\Upsilon$(2S) and $\Upsilon$(3S) from $\Upsilon(nS) \to \Upsilon(mS)\pi^+\pi^-$ followed by $\Upsilon(mS) \to \mu^+\mu^-$ but it is ineffective against cascade decays containing only neutral particles.
To reduce this background, we require fewer than two extra showers with more than 50 MeV (100 MeV) energy in the barrel (endcap) section of the calorimeter.
This requirement significantly suppresses the background while the direct muon efficiency decreases by less than 1\% (Fig.~\ref{fig:cascade}).
We estimate the remaining cascade background using measured branching fractions \cite{PDG} and a Monte Carlo simulation.
The residual cascade background is ($2.9\pm 1.5$)\% and ($2.2\pm 0.7$)\% for $\Upsilon$(2S) and $\Upsilon$(3S), respectively, where the uncertainty is dominated by the leptonic branching fractions and the selection efficiencies.

\begin{figure}
\includegraphics*[width=4.5in]{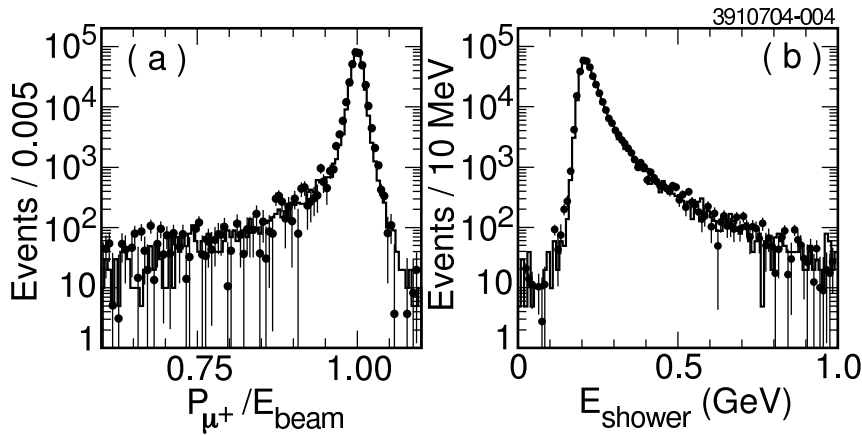}
\caption{Distribution of the scaled momentum (left) and the shower energy (right) of the $\mu^+$ candidates from $\Upsilon$(1S) decays after off-resonance subtraction (points) and from resonance Monte Carlo simulations (histogram). The vertical scale is logarithmic.}
\label{fig:data_mc}
\end{figure}

The overall selection efficiency for $\Upsilon \to \mu^+\mu^-$ decays is ($65.2 \pm 1.2$)\% from a GEANT-based \cite{GEANT} Monte Carlo simulation.
The relative systematic uncertainty in the efficiency is 1.8\% which is dominated by the uncertainty in the detector simulation (1.7\%) determined from a detailed comparison between data and Monte Carlo distributions of all selection variables (Fig.~\ref{fig:data_mc}).

The $\mu^+\mu^-$ selection has been checked by calculating the $e^+e^- \to \mu^+\mu^-$ cross section using the number of $\mu^+\mu^-$ events observed in the off-resonance samples, the corresponding Monte Carlo efficiency, and the integrated luminosity determined from Bhabha events.
The measured cross section is consistent with the theoretical cross section including higher order radiative corrections \cite{FPAIR}.

When selecting hadronic events, we minimize the systematic uncertainty by maintaining high selection efficiency.
QED backgrounds ($e^+e^- \to e^+e^-/\mu^+\mu^-/\gamma\gamma$) are suppressed by requiring $\geq3$ charged particles. 
In addition, for low multiplicity events with $<5$ charged tracks, we require the total energy detected in the electromagnetic calorimeter to be more than 15\% of $E_{\rm cm}$, and either the total calorimeter energy to be less than 75\% of $E_{\rm cm}$ or the most energetic shower to be less than 75\% of $E_{\rm beam}$.
To suppress beam-gas and beam-wall interactions we reject events in which the total energy visible in the calorimeter and in the tracking system is less than 20\% of $E_{\rm cm}$. 
We also use the event vertex position to reject the beam-related background as well as cosmic rays and to estimate the residual background from these sources.
Background to the hadrons from $\Upsilon \to \tau^+\tau^-$ decay is estimated to be ($0.7 \pm 0.1$)\%, ($0.4 \pm 0.3$)\%, and ($0.5\pm 0.2$)\% for the $\Upsilon$(1S), $\Upsilon$(2S), and $\Upsilon$(3S), respectively, where the uncertainty is dominated by the inaccuracy in the leptonic branching fractions and in the Monte Carlo efficiency.

We determine the selection efficiency for $\Upsilon \to hadrons$ from Monte Carlo simulation of the detector using event generators based on Jetset 7.3 and 7.4 \cite{Jetset74}.
The two simulations result in a slightly different efficiency and a comparison of the distributions of the selection variables in data and Monte Carlo simulations suggests that the real efficiency is apparently in between.
Hence, we average the two Monte Carlo efficiencies and assigned a relative systematic uncertainty to cover the difference between the two simulations: 1.6\% (1S), 1.2\% (2S) and 1.3\% (3S).
Uncertainties in the cascade and leptonic branching fractions of the $\Upsilon$ resonances contribute with an additional 0.3\% added in quadrature to the total efficiency uncertainty in the case of the $\Upsilon$(2S) and $\Upsilon$(3S).
The overall selection efficiency for hadronic resonance decays varies between 96-98\% (largest for the 1S and smallest for the 2S) depending on the relative rate of cascade decays that can produce a stiff $e^+e^-$ or $\mu^+\mu^-$ in the final state.

\begin{figure}
\includegraphics*[width=4.5in]{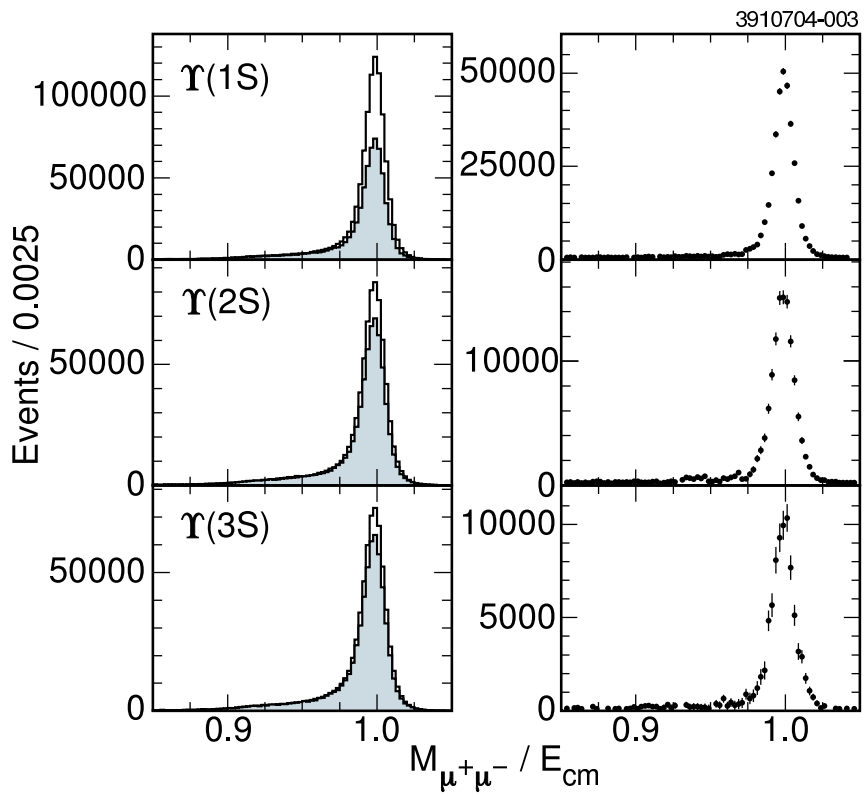}
\caption{Muon pair invariant mass distributions in on-resonance (empty) and scaled off-resonance (shaded) data on the left and the difference between these two distributions on the right. }
\label{fig:muons}
\end{figure}

Table~\ref{tab:result} presents the observed number of $\mu^+\mu^-$ and hadronic events from resonance decays ($\tilde{N}$), and the corresponding selection efficiencies ($\varepsilon$), along with their statistical uncertainties.
The invariant mass distribution of the $\mu^+\mu^-$ candidates in the on-resonance and off-resonance samples and after off-resonance subtraction are shown in Fig.~\ref{fig:muons}.

\begin{table}
  \centering
  \caption{Number of resonance decays to $\mu^+\mu^-$ and hadrons ($\tilde{N}$), selection efficiencies ($\varepsilon$), and the muonic branching fractions after correcting for interference. The uncertainty is statistical only.}\label{tab:result}
\bigskip
\begin{tabular}[c]{lccc}
  \hline
  \hline
    & $\Upsilon$(1S) & $\Upsilon$(2S) & $\Upsilon$(3S) \\
  \hline
   $\tilde{N}_{\mu\mu}$ ($10^3$) & $344.9 \pm 2.5$ & $119.6 \pm 1.8$ & $81.2 \pm 2.7$ \\
   $\varepsilon_{\mu\mu}$ & $0.652\pm0.002$ & $0.652\pm0.002$ & $0.652\pm0.002$ \\
   $\tilde{N}_{\rm had}$ ($10^6$) & $18.96 \pm 0.01$ & $7.84 \pm 0.01$ & $4.64 \pm 0.01$ \\
   $\varepsilon_{\rm had}$ & $0.979 \pm 0.001$ & $0.965 \pm 0.001$ & $0.975 \pm 0.001$ \\
  \hline
   ${\cal B}_{\mu\mu}$ (\%) & $2.49 \pm 0.02$ & $2.03 \pm 0.03$ & $2.39 \pm 0.07$ \\
  \hline
  \hline
\end{tabular}
\end{table}

Since the effect of interference between the resonant and non-resonant production is energy dependent (expected to be destructive (constructive) below (above) the resonance \cite{interference}), and its relative contribution to $\mu^+\mu^-$ is about ten times larger than to hadrons, the measured branching fraction depends slightly on the center-of-mass energy at which the data were taken.
We estimate the effect of interference at the luminosity-weighted average center-of-mass energies of the on-resonance as well as the off-resonance samples using a convolution of an interference-corrected Breit-Wigner resonance shape, a Gaussian energy spread, and a radiative tail \cite{Kuraev-Fadin}.
The resulting correction factors to the observed $\cal{B}_{\mu\mu}$ branching fractions due to interference are 0.984, 0.961, and 0.982 for the $\Upsilon$(1S), $\Upsilon$(2S), and $\Upsilon$(3S), respectively.
The $\Upsilon(nS) \to \mu^+\mu^-$ branching fractions listed in Table~\ref{tab:result} are corrected for interference.

The total fractional systematic uncertainties in ${\cal B}_{\mu\mu}$ are 2.7\% (1S), 3.7\% (2S), and 4.1\% (3S), respectively.
They are the quadrature sums of the fractional uncertainties due to several sources listed in Table \ref{tab:systematics}.
The systematic uncertainty in the selection efficiency ($\varepsilon$) is from detector modeling (dominant), trigger efficiency and Monte Carlo statistics.
The systematic uncertainty in the raw event number ($\tilde{N}$) is due to uncertainties in various backgrounds mentioned earlier.
Uncertainties in the interference calculation and variations in the center-of-mass energy contribute 1\%.
The dominant source of the systematic uncertainty in the cases of $\Upsilon$(2S) and $\Upsilon$(3S) is due to the uncertainty in the scale factor between the on-resonance and off-resonance data.

\begin{table}
  \centering
  \caption{Fractional systematic uncertainties (\%) in ${\cal B}_{\mu\mu}$.}
  \label{tab:systematics}
\bigskip
\begin{tabular}[c]{lp{0.3in}cp{0.3in}cp{0.3in}c}
  \hline
  \hline
   $\Upsilon$ && 1S && 2S && 3S \\
  \hline
   $\varepsilon_{\rm had}$ && 1.6 && 1.3 && 1.4 \\
   $\tilde{N}_{\rm had}$ && 0.2 && 0.3 && 0.4 \\
   $\varepsilon_{\mu\mu}$ && 1.8 && 1.8 && 1.8 \\
   $\tilde{N}_{\mu\mu}$ && 0.1 && 1.6 && 0.9 \\
   Scale factor && 0.8 && 2.3 && 3.1 \\
   Interference && 1.0 && 1.0 && 1.0 \\
  \hline
   Total && 2.7 && 3.7 && 4.1 \\
 \hline
 \hline
\end{tabular}
\end{table}

The final branching fractions, including systematic uncertainties, are
${\cal B}_{\mu\mu}(1S) = (2.49 \pm 0.02 \pm 0.07)$\%,
${\cal B}_{\mu\mu}(2S) = (2.03 \pm 0.03 \pm 0.08)$\%, and
${\cal B}_{\mu\mu}(3S) = (2.39 \pm 0.07 \pm 0.10)$\%.
The result for the $\Upsilon$(1S) is in very good agreement with the current world average of ($2.48 \pm 0.06$)\% \cite{PDG}, while our $\Upsilon$(2S) and $\Upsilon$(3S) results are about $3\sigma$ larger than the world average of ($1.31 \pm 0.21$)\% and ($1.81 \pm 0.17$)\% \cite{PDG}, respectively.

The total decay width of the resonances can be expressed as \cite{PDG}
\begin{equation}
\Gamma = \frac{\Gamma_{ee}\Gamma_{\rm had}/\Gamma}{{\cal B}_{\mu\mu}(1 - 3{\cal B}_{\mu\mu})}.
\end{equation}
Our improved muonic branching fractions, combined with the current values of $\Gamma_{ee}\Gamma_{\rm had}/\Gamma$ \cite{PDG} lead to the following new values for the total decay widths of the three narrow $\Upsilon$ resonances: $\Gamma (1S) = (52.8 \pm 1.8)$ keV, $\Gamma (2S) = (29.0 \pm 1.6)$ keV, and $\Gamma (3S) = (20.3 \pm 2.1)$ keV.
The uncertainties are the quadrature sums of the statistical and systematic uncertainties.

The new total widths of the $\Upsilon$(2S) and $\Upsilon$(3S) have a significant impact on the comparison between theoretical and experimental values of hadronic and radiative widths of these resonances since the experimental widths are determined as a product of the total widths and the measured transition branching fractions.
The new value of ${\cal B}_{\mu\mu}$(2S) also significantly lowers ${\cal B}(\Upsilon(3S) \to \pi\pi\Upsilon(2S))$ and ${\cal B}(\Upsilon(3S) \to \gamma\gamma\Upsilon(2S))$ (consequently ${\cal B}(\chi_b(2P_J) \to \gamma\Upsilon(2S))$ as well) when they are extracted from the measured exclusive $\Upsilon(3S) \to \pi\pi\ell^+\ell^-$ and $\gamma\gamma \ell^+\ell^-$ branching fractions.

In summary, we have measured the muonic branching fraction of the narrow $\Upsilon$ resonances below the open-beauty threshold with 2.8\%, 4.0\%, and 5.1\% relative uncertainty.
The obtained branching fractions for the $\Upsilon$(2S) and $\Upsilon$(3S) resonances are significantly larger than prior measurements and the current world average values, resulting in narrower total decay widths.
The new branching fractions, particularly ${\cal B}_{\mu\mu}(2S)$, also affect the measured rates of other transitions leading to the $\Upsilon$ resonances and observed by the subsequent decay $\Upsilon \to \mu^+\mu^-$.

We gratefully acknowledge the effort of the CESR staff 
in providing us with excellent luminosity and running conditions.
This work was supported by the National Science Foundation
and the U.S. Department of Energy.

\end{document}